\documentclass[aps,prb,showpacs,twocolumn,superscriptaddress]{revtex4}
\usepackage{graphicx}
\usepackage{dcolumn}
\usepackage{bm}
\usepackage{color}
\usepackage{times}

\begin{document}

\title{Time-Resolved X-ray Microscopy of Spin-Torque-Induced Magnetic Vortex Gyration}

\author{Markus Bolte}
\email{mbolte@physik.uni-hamburg.de}
\author{Guido Meier}
\affiliation{Institut f\"ur Angewandte Physik und Zentrum f\"ur
Mikrostrukturforschung, Universit\"at  Hamburg, Jungiusstrasse 11,
20355 Hamburg, Germany}

\author{Benjamin Kr\"uger}
\affiliation{I. Institut f\"ur Theoretische Physik, Universit\"at
Hamburg, Jungiusstrasse 9, 20355 Hamburg, Germany}

\author{Andr\'{e} Drews}
\author{Ren\'{e} Eiselt}
\author{Lars Bocklage}
\affiliation{Institut f\"ur Angewandte Physik und Zentrum f\"ur
Mikrostrukturforschung, Universit\"at  Hamburg, Jungiusstrasse 11,
20355 Hamburg, Germany}

\author{Stellan Bohlens}
\affiliation{I. Institut f\"ur Theoretische Physik, Universit\"at
Hamburg, Jungiusstrasse 9, 20355 Hamburg, Germany}

\author{Tolek Tyliszczak}
\affiliation{Advanced Light Source, LBNL, 94720 Berkeley,
California, USA}

\author{Arne Vansteenkiste}
\affiliation{Department of Subatomic and Radiation Physics, Ghent
University, Proeftuinstraat 86, 9000 Ghent, Belgium}

\author{Bartel Van Waeyenberge}
\affiliation{Department of Subatomic and Radiation Physics, Ghent
University, Proeftuinstraat 86, 9000 Ghent, Belgium}
\affiliation{Max Planck Institut f\"ur Metallforschung,
Heisenbergstrasse 3, 70569 Stuttgart, Germany}

\author{Kang Wei Chou}
\author{Aleksandar Puzic}
\author{Hermann Stoll}
\affiliation{Max Planck Institut f\"ur Metallforschung,
Heisenbergstrasse 3, 70569 Stuttgart, Germany}

\date{30 January 2008}

\begin{abstract}
Time-resolved X-ray microscopy is used to image the influence of alternating high-density currents on the magnetization dynamics of ferromagnetic vortices. Spin-torque induced vortex gyration is observed in micrometer-sized permalloy squares. 
The phases of the gyration in structures with different chirality are compared to an analytical model and micromagnetic simulations, considering both alternating spin-polarized currents and the current's Oersted field. In our case 
the driving force due to spin-transfer torque is about $70 \%$ of the total excitation while the remainder originates from the current's Oersted field. This finding has implications to magnetic storage devices using spin-torque driven magnetization switching and domain-wall motion.
\end{abstract}


\maketitle 
The discovery that spin-polarized electrons traveling through ferromagnets apply a torque on the local magne\-ti\-za\-tion \cite{BergerJAP1984} opened up a new field of research in solid state physics that could potentially result in new magnetic storage media. 
It is now understood that the spin-transfer torque acts on inhomogeneities in the magnetization, e.g., on interfaces between magnetic layers,\cite{Slonczewski1996} on domain walls,\cite{ZhangLiPRL2004, BenjaminPRB2007} i.e., interfaces between regions of uniform magnetization, or on magnetic vortices.\cite{ShibataPRB2006, KasaiPRL2006, BenjaminVortexOscillator2007, YamadaNatureMat2007} Magnetic domain walls, usually vortex walls,\cite{ThiavilleJMMM2005} can be driven by spin-polarized currents to store information in bit registers.\cite{ParkinPatent2004}

Vortices appear in laterally confined thin films when it is energetically favorable for the magnetization to point in-plane and parallel to the edges. In the center the magne\-ti\-za\-tion is forced out-of-plane to avoid large angles between magnetic moments that would drastically increase the exchange energy. The region with a strong out-of-plane magnetization component is called the vortex core and is only a few nanometers in diameter.\cite{ShinjoScience2000, WachowiakScience2002} The direction of the magnetization in the vortex core, also called the {\em core polarization p}, can only point out-of- or into-the-plane ($p$=$+1$ or $p$=$-1$, respectively). Hence ferromagnetic thin films containing vortex cores have been suggested as data storage elements. The {\em chirality} $c = +1 (-1)$ denotes the counterclockwise (clockwise) in-plane curling direction of the magnetization. It is known that vortices can be excited to gyrate around their equilibrium position by magnetic fields.\cite{ArgylePRL1984, ChoeScience2004} 
Recently it has been shown that field excitation can also switch the core polarization.\cite{BartelNature2006, HertelPRL2007, ShekaAPL2007, KimAPL2007, LeePRB2007, XiaoJAP2007} Micromagnetic simulations predict that spin-polarized currents can cause vortices both to gyrate\cite{ShibataPRB2006, BenjaminVortexOscillator2007} and to switch their polarization.\cite{YamadaNatureMat2007, CaputoPRL2007, LiuAPL2007} Both for field- and spin-torque-driven excitation, the direction of gyration is governed by the vortex polarization according to the right-hand rule (see Fig. 2 of Ref.\cite{ChoeScience2004}). The phase of field-driven gyration depends also on the chirality, while spin-torque driven gyration is independent of the chirality as the spin-transfer torque is proportional to the spatial derivative of the magnetization.\cite{BenjaminVortexOscillator2007} Time- and spatially averaging experimental techniques indicate that spin-torque-driven vortex gyration and switching indeed occurs, but conclusive evidence by time-resolved domain imaging technique that resolves the phase of gyration is still elusive.

%
Here we show by time-resolved X-ray microscopy that magne\-tic vortices in confined structures can be excited to gyration by high-frequency currents of high density passing directly through the ferromagnetic element. 
By observing the phase of the gyration relative to the excitation, we can discriminate between the current's spin-torque and its Oersted field con\-tri\-bu\-tions to the vortex motion. Field strengths of $30~\mu$T that are due to the current in the gold contacts and an inhomogeneous current distribution within the ferromagnetic element itself are calculated.

\begin {figure} [!t]
\includegraphics[width=8 cm] {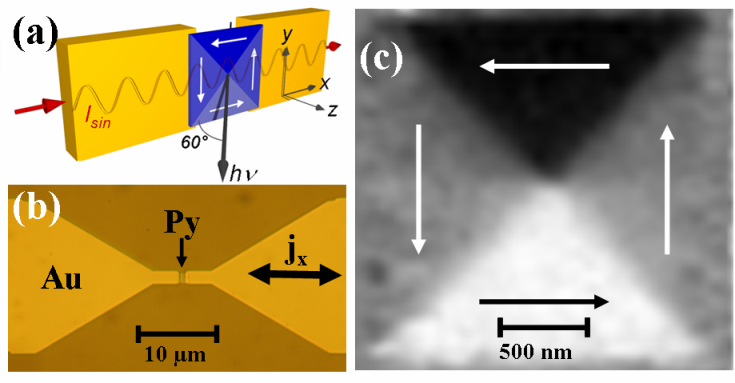}
\caption{\label{Fig1}(color online) (a) Scheme of the permalloy square contacted by two gold wires. The sample is tilted by 60$^{\circ}$ relative to the incident X-ray beam. (b) Optical micrograph of a permalloy square and its contacts on the Si$_3$N$_4$-membrane. (c) Magnetic contrast of the relaxed permalloy square with a thickness of 20~nm showing the $x$-component of the magnetization as black-to-white contrast.}
\end{figure}

We investigated 2~$\times$~2~$\mu$m$^2$ large and 20~nm thick permalloy (Ni$_{80}$Fe$_{20}$) squares in which Landau-domain patterns with a single vortex are energetically favorable at remanence. Microstructured permalloy squares were prepared on $200$~nm thin Si$_3$N$_4$ membranes for minimal absorption of the X-rays. The squares were prepared onto the membranes by electron-beam lithography, electron-beam evaporation, and lift-off processing. To excite the structures with alternating currents, they were contacted by $40$~nm thick gold strip lines with an overlap of $150$~nm as shown in Fig.~\ref{Fig1}. Thus the current had to pass through the ferromagnetic material. Additional permalloy squares having the same dimensions were placed completely underneath the strip lines to compare the phases of the current-driven gyration to field-driven gyration within the same experiment. 

The magnetization was excited by high density AC-currents, and the magnetization's response was imaged by time-resolved X-ray microscopy. For this the samples were placed in the scanning transmission X-ray microscope (STXM) of beam line 11.0.2 at the Advanced Light Source (ALS) in Berkeley. 
The monochromatic, circularly polarized X-ray beam from the undulator beam line was focused onto the sample with the help of a Fresnel zone plate. The resolution of this X-ray microscope was about $30$~nm. 
The sample was scanned in the $xy$-plane with a high resolution scanning stage under interferometric control and the transmitted intensity was recorded. The photon energy was set at the Ni L$_3$-absorption edge (852.7 eV), where X-ray circular dichroism (XMCD) \cite{SchützPRL1987} yields the magnetic contrast. With XMCD, the transmitted photon intensity is higher when the magnetic moments and polarization are antiparallel than in the parallel case. In our case, the sample plane was set at an angle of 60$^\circ$ with respect to the incident beam (see Fig. \ref{Fig1}(a)) so that the microscope can detect the in-plane magnetization. 
We can thus unambiguously determine the chirality of the vortices. 

\begin {figure}
\includegraphics[width=8 cm] {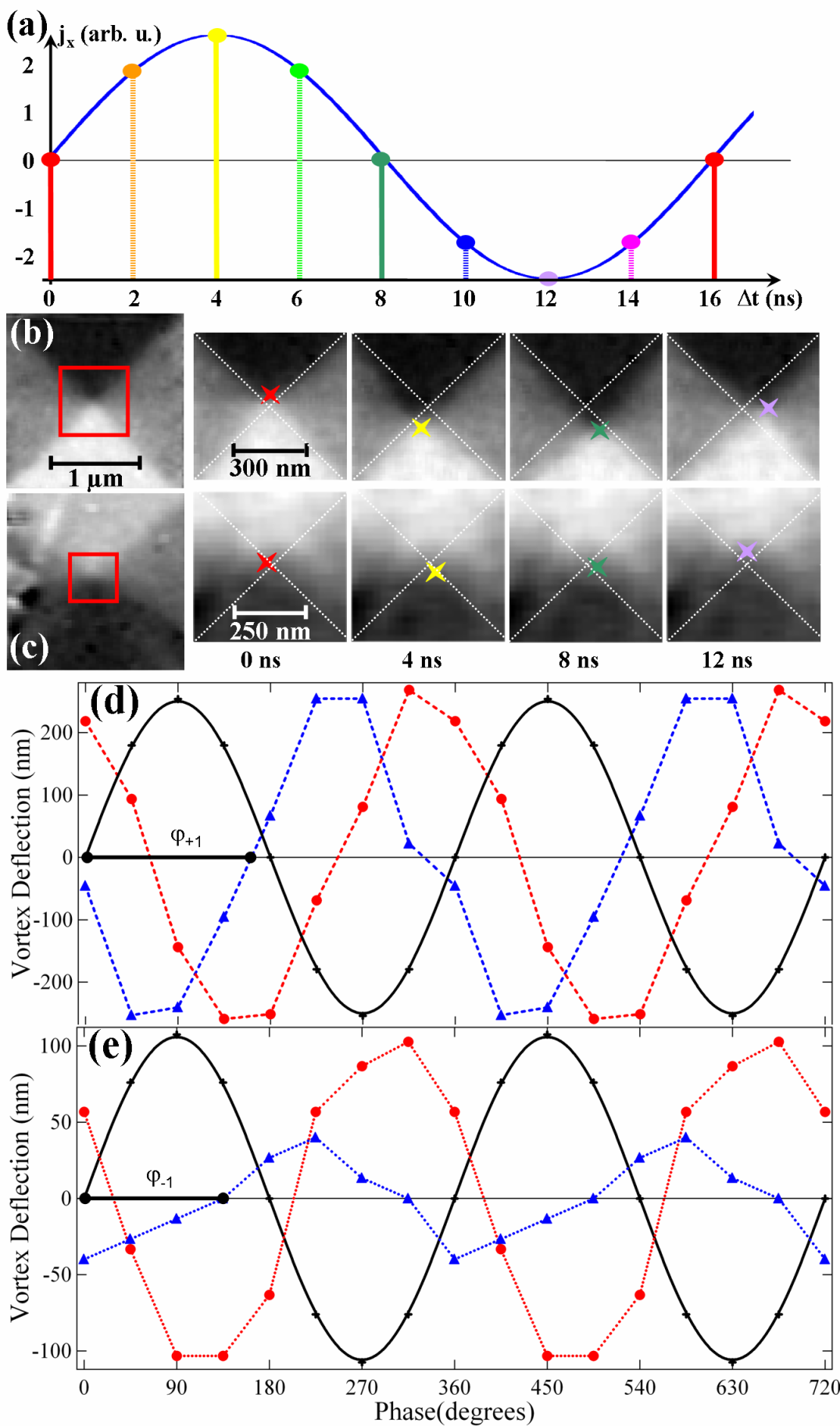}
\caption{\label{Fig2}(color online) (a) Sampling of the response to an 62.5 MHz AC-current excitation (j$_x$) at eight different phases 
in steps of 45$^\circ$. Shown in (b) and (c) are X-ray images of complete Landau-domain pattern (left) and blow-ups of the center piece at four channels corresponding to phases $0^\circ$, $90^\circ$, $180^\circ$, and $270^\circ$ (right). 
(d) and (e) show the vortex deflection in $x$ direction (blue) and $y$ direction (red) depicted in (b) and (c), respectively. The black solid curves represent the exciting current. Points represent the measured data, lines are guides to the eye.}
\end{figure}

The temporal resolution of the microscope, here about $70$~ps, is given by the width of the electron bunches that produce the X-ray photon flashes. In the standard multi-bunch operation mode of the synchrotron used here, the flash repetition rate is $500$~MHz. To resolve the individual flashes, a fast avalanche photo diode was used as a photon detector. With fast data acquisition electronics, the signals from individual bunches were recorded.\cite{AcremannRSI2007} At the ALS, one of the 328 electron bunches has a much larger amplitude. It produces a brighter flash and is used as a reference marker to align the excitation signal with the data acquisition. The absolute phase relation between the recorded images and the excitation current is made by sending a short pulse through the detector electronics. By aligning its arrival to the pulse produced by the photons of the reference marker an accuracy 
of approximately $100$~ps is achieved. A signal generator was synchronized with the X-ray flashes of the ALS, and an excitation frequency of $500$~MHz$/8=62.5$~MHz was selected since it is close to the expected resonance frequency of the vortex.\cite{NovosadPRB2005} 
The alternating current was sent through the permalloy squares and the magnetic response was detected at different phases (see Fig.~\ref{Fig2}(a)). Figures ~\ref{Fig2}(b) and ~\ref{Fig2}(c) show the magnetic contrast of two different samples at eight different phases of the excitation.\cite{footnote1} 

\begin {figure}
\includegraphics[width=8 cm] {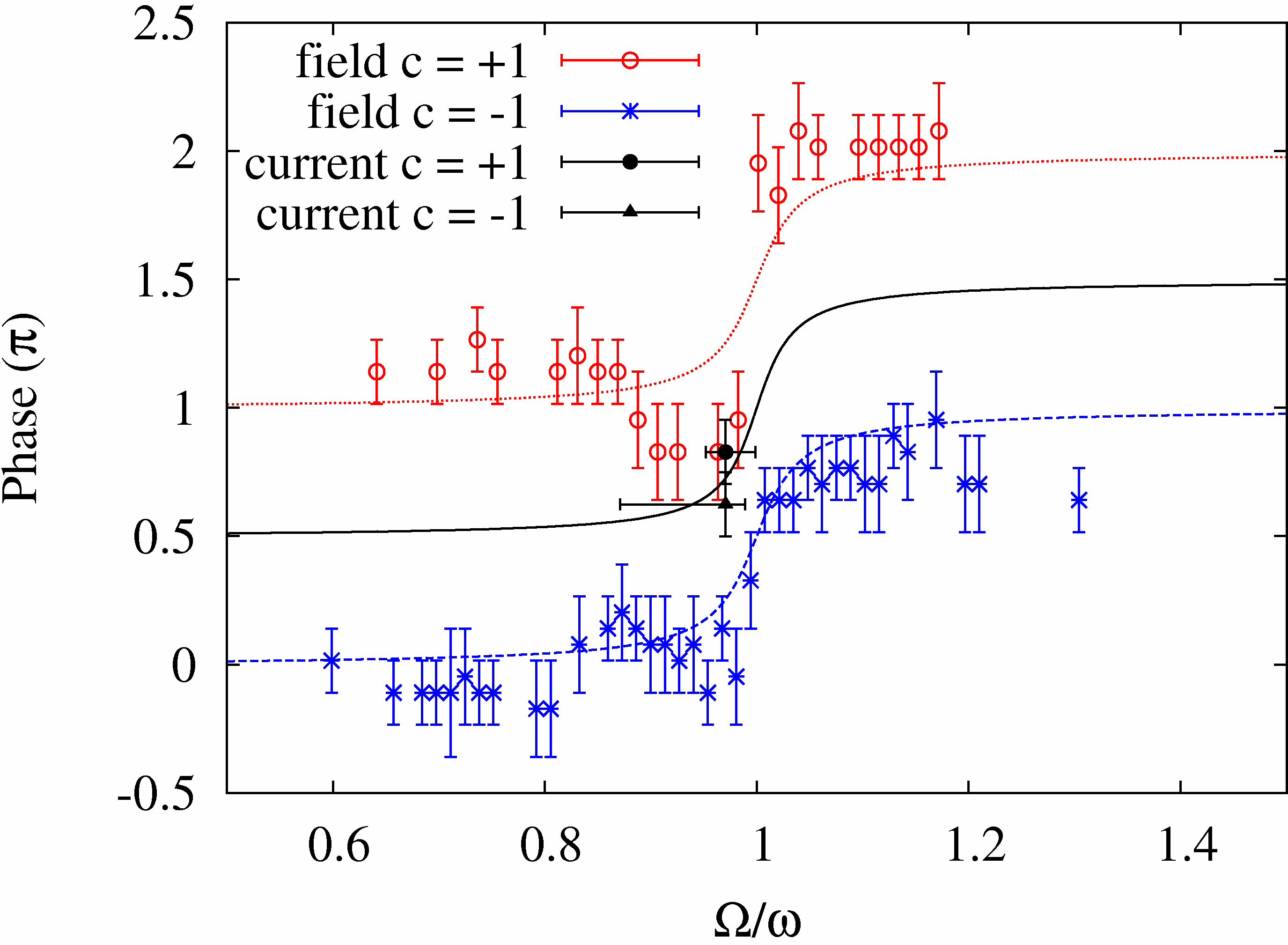}
\caption{\label{Fig3} (color online) Phase response of magnetic vortex oscillators as detected by time-resolved X-ray microscopy normalized to the respective resonance frequency. The red circles and the blue stars represent the phase responses of two permalloy squares excited purely by an AC-magnetic field. The black circle and triangle represent the phases of the current-fluxed samples shown in Fig. \ref{Fig2}(b) and (c), respectively.}
\end{figure}

The permalloy square ($c=+1$) in Fig.~\ref{Fig2}(b) was excited with a current density amplitude of $j=1.2\cdot10^{11}$~A$/$m$^2$. The vortex performs a counterclockwise gyration, it must therefore have a positive polarization ($p=+1$).\cite{ChoeScience2004} The amplitude of the gyration is $250\pm20$~nm, i.e., the vortex gyrates at a velocity of $100\pm8$~m$/$s. The vortex in Fig.~\ref{Fig2}(c) was excited with a lower excitation amplitude of $j=4.7\cdot10^{10}$~A$/$m$^2$. It has a negative chirality ($c=-1$) and also gyrates counterclockwise, i.e., $p=+1$. As can be expected, the amplitude of gyration is much smaller. 
From the analysis of the vortex position at certain excitation phases and by using differential images,\cite{PuzicJAP2005} the relative phase with respect to the excitation can be derived (see Fig. \ref{Fig2}(d) and (e)). 
The phase difference $\Delta\varphi=\varphi_{+1}-\varphi_{-1}$ between the gyration of the vortices with different chiralities is about $45^{\circ}$. The permalloy squares underneath the strip lines were also imaged and analyzed in like manner. They showed phase differences of $180^{\circ}$ for vortices with opposite chirality as shown in Fig. \ref{Fig3}.

To better understand the dependence of the phases on current and field excitation, micromagnetic simulations were conducted. AC-excitations of a small permalloy square with either spin-polarized currents ${j_x}$ or magnetic fields ${H_y}$ were simulated at different frequencies, i.e., below resonance, at resonance, and above resonance, for all chiralities and polarizations. For the simulations the Object Oriented Micromagnetic Framework ({\em OOMMF}) was extended by additional spin-torque terms.\cite{ZhangLiPRL2004, BenjaminPRB2007} A Landau-domain pattern of $200\times200\times20~$nm$^{3}$ with a vortex was chosen.\cite{footnote2}
We assumed a saturation magnetization of $M_{\mathrm s}=8\cdot 10^{5}$~{A/m}, an exchange constant of $A=1.3\cdot 10^{-11}$~{J/m}, a Gilbert damping of $\alpha=0.01$, and a ratio $\xi= \alpha$ between spin-flip and spin-relaxation time.\cite{HayashiPRL2007,GuidoPRL2007} The amplitude of the spin-polarized current was set to $P\cdot j=2.5\cdot 10^{10}${A/m}, and the amplitude of the magnetic field was $H=250$~A/m$^2$. The size of the simulation cells was $2\times 2\times 20~$nm$^{3}$. The resonance frequency $\omega_{r}$ of the vortex element was derived by fitting the motion of a vortex relaxed from an initially excited state to the equation of motion of a damped harmonic oscillator.\cite{BenjaminVortexOscillator2007} The simulations show that the phase with respect to the AC-excitation differs between spin-transfer-torque- and field-excitation, because the latter depends on the chirality, in agreement with previous micromagnetic simulations.\cite{ShibataPRB2006, KasaiPRL2006} 

Following the 'rigid model' for magnetic vortices in thin films,\cite{ThielePRL1973, HuberPRB1982} the vortex gyration due to alternating fields can be described by a two-dimensional harmonic oscillator.\cite{LeeCondmat2007} This model can be generalized to include torques due to spin-polarized currents.\cite{LeeAPL2007, BenjaminVortexOscillator2007} In this model, the alternating magnetic field or current forces the vortex to oscillate along the direction of the field or current. The magnetostatic field created by the deviation of the vortex from its equilibrium position drives the oscillator perpendicular to the excitation. For a current passing through the permalloy square in $x$ direction (see Fig.~\ref{Fig1}), the resulting Oersted field is in $y$ direction. Thus the solution of the equation of motion of the vortex core in the oscillator model can be written as\cite{BenjaminVortexOscillator2007}
\begin {equation} \begin{array}{cl}
\label{eqn1} \left(\begin{array}{c} X\\Y\end{array}\right)& =
\frac{e^{i \Omega t} } {\omega^2 + \left( i \Omega + \Gamma \right)^2 } \\& \times \left( \frac{\gamma l } {2 \pi} c H_y \left(\begin{array}{c} -\omega \\i \Omega p\end{array} \right) - b_j j_x  \left(\begin{array}{c} i \Omega  \\
p \omega \end{array} \right) \right). \end{array}
\end {equation}

Here, $\Omega$ is the excitation frequency, $\omega$ the gyration frequency of the free vortex, $\Gamma$ the damping constant of the vortex harmonic oscillator which is proportional to the Gilbert damping $\alpha$, and ${\gamma = 2.21 \cdot 10^5 \mbox{ m/As}}$ the gyromagnetic ratio for permalloy. The constant ${b_j=P\mu_{\mathrm B}/(eM_{\mathrm s})}$ with the saturation magnetization $M_s$ and the spin polarization $P$ describes the coupling between the electrical current and the magnetization. 

The model is used to extract the contributions of Oersted field and spin-torque to the phases observed in the experimental data. 
The blue crosses shown in Fig.\ref{Fig3} represent the phases of field-driven vortex-core gyration ($c=-1$) with respect to the excitation at different excitation frequencies, while the dashed blue line is a fit from Eq. \ref{eqn1}. The red circles and the corresponding fit (dotted red line) are from a vortex having $c=+1$.\cite{footnote3} The good agreement between the experimental data and the fit shows the validity of the harmonic oscillator model. The X-ray microscopy data from the current-fluxed samples is plotted as a black circle and a black triangle, and the corresponding fit is shown by a solid black line. The error bars in $y$ direction are derived by error propagation from the measurements. The error bars in $x$ direction follow from an uncertainty in the resonance frequency when assuming the resonance frequencies of the vortices are not the same. 
By estimating the Oersted field due to the current using Eq. 1 and $\Delta\varphi=45^{\circ}$, 
values of up to 30 $\mu$T for a current density of $1.2 \cdot 10^{11} $A$/$m$^2$ are derived. The driving force on the vortex due to an Oersted field of this magnitude corresponds to about $30\%$ of the total driving force. Micromagnetic simulations using {\em both} spin-torque and field excitation with current and field values as calculated above yielded 40$^{\circ}$ phase difference between vortices having $c=+1$ and $c=-1$. This is almost the same phase difference as observed by X-ray microscopy. 

Three-dimensional current-path and Oersted-field calculations were also performed for our sample geometry, taking an inhomogeneous current density in the gold contacts and in the permalloy into account. The calculations showed fields of very similar magnitude ($20~\mu$T) as deduced from the experiments. The fields originate from the perpendicular currents leading from the contacts into the permalloy as well as from an inhomogeneous current density in the permalloy, because the current enters and exits the permalloy on the upper side of the ferromagnet. For smaller and thicker vortex geometries, the influence of the current leads and the inhomogeneous current density is much larger. For example, permalloy layers of $1~\mu$m length and $50$ nm thickness with $40$ nm thick gold contacts and higher current densities than in our cases lead to Oersted-field strengths of up to several Millitesla, more than sufficient for field-induced switching. 

The presence of Oersted fields due to an asymmetric setup can be seen as a general challenge to spin-torque experiments, but so far have not been taken into account. One must be careful to rule out field-driven or field-assisted magnetization dynamics in spin-torque experiments. In our case the observations of the phase of magnetic vortex gyration were possible due to its periodic motion in a confined structure. The dynamics of other magnetic objects, e.g., vortex-domain walls, are more difficult to record, but are also subject to Oersted fields from spin-polarized currents. As seen from our observations, one cannot safely assume that the change of a magnetic structure is due to a traversing spin-polarized current alone.

With time-resolved X-ray microscopy we have observed magnetic vortex gyration driven by spin-polarized currents that can be described by a harmonic oscillator model. We identified the spin torque as the main driving force, however, we have also recognized a non-negligible contribution of the current's Oersted field. 
In experiments, one needs to resolve the phase and the sense of gyration to separate the contribution of the current's spin-torque to magnetic vortex gyrations from the current's Oersted field. 
These observations are relevant to technological applications since spin-polarized currents that switch the polarization of vortices have been suggested for data storage devices \cite{YamadaNatureMat2007}. 

\begin{acknowledgments}
We thank Ulrich Merkt for valuable discussions. Financial support from the Deutsche For\-schungs\-gemein\-schaft via the SFB 668 "Magnetismus vom Einzelatom zur Nanostruktur" as well as the GK 1286 "Functional Metal-Semiconductor Hybrid Systems" is gratefully acknowledged.
\end{acknowledgments}

\end{document}